\documentclass[9pt,a4paper]{extarticle}
\pdfoutput=1
\usepackage[pdftex]{graphicx}                        
\usepackage[paper=a4paper, margin={1.75cm},centering]{geometry}
\usepackage{mathpazo}
\usepackage{inconsolata}                             
\usepackage[T1]{fontenc}
\usepackage[utf8]{inputenc}
\usepackage[hyphens,obeyspaces,spaces]{url}
\usepackage{xcolor}
\usepackage{multicol}

\graphicspath{{figures/}}                           
\date{}

\def\thisday{\the\year-\ifcase\month\or Jan\or Feb\or Mar\or Apr\or May\or Jun\or Jul\or Aug\or Sep\or Oct\or Nov\or Dec\fi -{\ifnum \day<10 0\fi}\number\day} 

\def\citenum#1{\def\@cite##1##2{##1}\cite{#1}}

\newcommand{\mycomment}[1]{}
\definecolor{grey}{rgb}{0.85,0.85,0.85}
\definecolor{middlegreen}{rgb}{0,0.6,0}

\makeatletter %
   \renewcommand{\@biblabel}[1]{\textbf{\textsf{#1}.}}    
\makeatother

\begin{document}

\begin{center}
   \Huge\bf The problem of scientific greatness and the role of ordinary scientists \\[-5pt] {\normalsize Nathan Hagen, \thisday}
\end{center}

\vspace{15mm}
\setlength\columnsep{20pt}

\begin{center}
\begin{minipage}{0.75\linewidth}
   \textbf{Abstract:} Textbooks in physics use science history to humanize the subject and motivate students for learning, but they deal exclusively with the heroes of the field and ignore the vast majority of scientists who have not found their way into history. What is the role of these invisible scientists --- are they merely the worker ants in the colony of science, whose main utility is to facilitate the heroes of the field?
\end{minipage}
\end{center}

\begin{multicols}{2}

\section{Introduction}

As students of the sciences, we start our studies filled with enthusiasm and ambition, drinking up technical details while reading about the theories and experiments of our great predecessors --- Newton, Darwin, Einstein, \textit{et al}. After years of study, as we gain in ability and reach the frontiers of knowledge in our chosen disciplines, we encounter the harsh realities of research and struggle to push forward. While some succeed and receive the accolades of their peers, many give up and leave science altogether, some tire of the struggle and focus on teaching, while others push on without wide recognition of their work. These latter two groups form the majority of practicing scientists --- the 99\% whose work rarely finds a place in the standard textbooks and yet who are nevertheless responsible for the majority of scientific publications. How are these to see their role in the whole effort? Are they really, as the disregard of most science biographies seems to imply, merely worker ants in the colony of science, whose main utility is to facilitate the work of our betters?

While this tension between ambition and reality is hardly unique to the sciences, it is something that most scientific researchers face, even if only subconsciously. Physicist Koichi Mano once wrote to his former teacher Richard Feynman of his feelings of inadequacy, to which Feynman replied by explaining how most of what the famous scientist had worked on was in fact second-rate work or outright failure: ``I have worked on innumerable problems that you would call humble, but which I enjoyed and felt very good about because I sometimes could partially succeed....  No problem is too small or too trivial if we can really do something about it.''~[\citenum{Feynman2008}, p.~201.] This feeling of progress and achievement, even if on only small problems, is important to the psychology of all creative work. In his reply, Feynman continues by explaining that the ideals imbibed by students --- the ideals of the heroic model of science --- are impractical for day-to-day research work. Even some scientists considered great by their contemporaries have suffered from feelings of inadequacy, as Paul Ehrenfest did in the 1930s.\cite{Johnson2019-Ehrenfest}

Framing a dramatic narrative for the biographies of famous scientists has been central to creating these impractical ideals. In these stories, a successful scientist's life is typically depicted as a hero's quest, and the vanquishing of the unknown as the conquest. While this convenient framework motivates each biography and drives its narrative, heroes of history are often figureheads for the armies that they lead. Without the armies to support them, they can achieve little of lasting value. And so it is with science, though scientific biographers work hard to draw a contrast between the primary subjects of their study and the often nameless group of colleagues surrounding them. While recognition of these colleagues' names has not survived, their efforts have nevertheless been essential for the advances achieved by science.

\section{What makes a scientist great?}

Some geniuses stand so far above us in their insights that they seem alien in their ability. Isaac Newton formulated the laws of motion, discovered a mathematical description of gravity, created calculus, invented the first practical reflecting telescope, built the theory of color based on the light spectrum, and added a host of other scientific accomplishments. Any one of these would have immortalized his name in science, and yet he did them all. Einstein discovered the photoelectric effect and thus the first convincing proof of the existence of photons as quanta; he derived the equivalence formula for mass and energy, saw that the Lorentz contraction implied the relativity of frames of reference, and realized that inertial motion having no preferential frame further implied a change to the laws of gravity, among many more accomplishments. The achievements of these two great scientists, however, stand so high that even to the scientific elite of today their accomplishments appear incredible. What historical narratives often fail to capture, however, is that these accomplishments are built upon the prior work of other (often forgotten) scientists, and that even the greatest achievements of the scientific elite merely move the timeline of science forward a little. If Newton had succumbed to sickness as a child, then it is difficult to imagine that the science of today would be much retarded by the change. His discoveries would have remained hidden by Nature for perhaps a few decades, but would have been discovered nevertheless. Almost nothing in science has the sense of being without a direct antecedent.

The two characteristics that we generally associate most with great scientists are their productivity for creating useful ideas and the profundity of those ideas. Two additional characteristics that go largely unmentioned, however, is their outstanding ability to communicate, and that many of them have been \emph{lucky}. While both of these latter characteristics are not what we usually associate with scientific greatness, they are nevertheless critical for achieving the highest status and recognition.

Researchers with greater skill in writing and speaking have a huge advantage over their less fluent peers in garnering recognition for their work. Clear explanations and proper emphasis makes it easier for others to understand new ideas. Moreover, announcing a discovery to the right people, and in the right way, brings rapid acceptance, whereas publishing a result in turgid prose within an obscure journal is little better than dropping the results into a trash bin. Elite scientists do a recognizably better job at communicating their work. Richard Feynman is widely respected as much for his non-technical writing and his teaching as for his scientific discoveries. For what is probably his most famous discovery --- the Feynman diagrams --- Feynman spent years talking to people, explaining, and basically proselytizing to get his idea recognized and understood. [\cite{Mlodinow2011}, p.29] Einstein's scientific papers are written with an unrivalled crystalline clarity and an apparent effortlessness that makes his ideas easy to follow. These communication skills are not just that of narrating a theory but rather of framing a story in which the theory is a part, in order to shape our understanding of its physical significance and importance.

The second little-acknowledged characteristic of the great scientists is that they were generally lucky.\cite{Pluchino2018} This may take the form of being born into a family of means for supporting their higher education, but also of having parents who foster scientific thinking from an early age. Later in their careers, luck may take the form of having just the right timing, of selecting just the right mentor, or of choosing just the right research project, to achieve results of fundamental importance.\cite{Zuckerman1996,Giaever2017} Physicists Lawrence Bragg and Brian Josephson have their names attached to discoveries made while in their early 20s, while pursuing research under the guidance of a senior scientist. While these examples may seem to indicate a lucky choice of direction for research, many people feel that even if a great scientist had not made the specific discoveries for which he or she later became famous, he or she would have worked on something else and have found something important there instead. That is, they feel that their greatness is \emph{robust}. But if we accept that luck is at least partly a factor, and that many of the great scientists are really only known for one great accomplishment, then the robustness of their greatness feels less convincing. Thomas Bayes, for example, was respected by his peers but is now known only for his work on developing one equation, and that published posthumously, with the rest of his life's work having no significant influence on modern science.

The advantages of luck and skillful communication, while not belonging to the technical part of scientists' discoveries, thus play a critical role in immortalizing many scientists' work and in deciding who receives the primary recognition for the achievement. Whether fair or not, scientists hampered by poor luck and poor communication skills are often left unrecognized.\cite{Arteaga2018} So much of scientific historical writing, including that written by scientists themselves, ignores these factors and builds a mythology founded on its heroes. This mythologizing, however, obscures the truth that the great scientific advancements were shaped by luck and writing skill, and were assisted by the efforts of forgotten researchers. The scientific elite themselves all began their careers in this category, as unrecognized researchers slowly working their way up through the hierarchy with a succession of accomplishments. Thus, there are many scientists who may be boosted into the ranks of the great with the award of a great prize, or the making of a single important new discovery. These often require being lucky, or being friendly with the right people. Without the extra lift provided by these symbols of recognition, many scientists would be categorized as ``merely successful'' rather than great. Max Planck, for example, was highly regarded among his peers around the end of the 19th century, but it is hard to believe that his name would be much recognized today if he had not decided to attack the problem of blackbody radiation. Although his solution is testament to his brilliance, we can also recognize in it a strong element of chance --- the timeliness of his choice and of his being so well prepared in the theoretical tools needed to find the quantum solution.

\section{The great man theory}

The ``great man theory'' is an idea that has been around throughout the ages but was promulgated in the 19th century and still influences historical writing and thinking today. Its central thesis is that there are men and women whose personal abilities (intelligence, charisma, etc.) are so far above the ordinary that they have a widespread and lasting influence over events. Accepting the great man theory, however, implies that one must also take seriously the idea that there could also be a corresponding ``negative great man''. This does not mean an insignificant man, but rather someone who is highly influential in a negative way, by impeding progress.\cite{Scott2018} There are more than a few physicists who view Niels Bohr's influence on quantum mechanics in this vein.\cite{Bell1993,Norsen2017} In the late 19th century, an unsympathetic view may put Lord Kelvin's evangelizing for the vortex atom theory, despite its lack of empirical validation, in a similar category. These are physics' equivalent of the incompetent monarchs in Europe's early modern period --- kings and queens whose decisions were more ruinous to their kingdoms than no decision at all.

Although influential, the ``great man'' concept is not without a countertheory. The obverse does not have a catchy name, but can be called the social constructivist theory --- that influential figures are the products of their environments and societies, so that their success is contingent on the social conditions around them. In other words, they were lucky, and they had help. The biographies of most scientists can be re-framed in this context --- that it is by their teachers and by the friendship of their peers that great scientists learned to focus their thoughts and achieve new insights. This perspective has influenced modern historiography to shift narratives away from individual leaders to social frameworks that describe the wider forces that shape history beyond those of single individuals. Thus, modern historical writing emphasizes that Alexander the Great would not have achieved such a wide influence over history if his father had not produced an army for him that was primed for conquest. Likewise, Caesar had the stolid Roman legions, Napoleon the {\'e}lan of revolutionary France's soldiery. 
In each case, these leaders would not have achieved ``greatness'' without their situation being prepared for them. They were not just unusually smart, they were also lucky. Similarly, some scientists arrive at the scene with just the right set of ideas to achieve impressive results.

Conflicting opinions about these two theories drives much discussion, but while the great man and the social constructivist theories are often phrased as being incompatible with one another, the truth clearly lies somewhere between. Men and women of outstanding genius clearly exist, but it is also clear that genius alone is not enough to lead them to extraordinary achievements. For that, they need luck and the help of their peers.

\section{Scientific discoveries as the tip of a pyramid of achievements}

Since most scientists are ignored by history, should we consider them just as worker-ants and an unfortunate waste of resources? This kind of condescending perspective follows the natural hierarchy in which many scientists prefer to view scientific endeavors, placing theoretical work at the pinnacle, experimental work filling out the middle, engineering at the bottom, and, one supposes, the nonscientific population trudging through the dirty unlit caverns underpinning the whole structure. This view persists even though engineering problems have inspired many of the greatest scientists to their highest accomplishments.


Many discoveries exist in the gray region between the great and the merely useful --- discoveries that have had a substantial impact but which have not been widely recognized. One example might be that of the Fourier transform spectrometer (FTS), which helped many fields of science to take measurements that were difficult or impossible with existing instrumentation. It also provided the first strong evidence for the advantages of computational sensing, a field that has expanded greatly with the increase in computing power over the past decades. However, since the invention of the FTS was a collective effort by a number of individuals, most of whom were not attention-seeking, no single researcher's name has become associated with it, and this may contribute to its unexpected obscurity. In modern science, most scientific achievements are the product of collective effort by many scientists that accumulate into a single recognized fundamental advance. Once the achievement is clearly demonstrated and acknowledged, the background work needed to get there is no longer useful and disappears from memory. Planck's great discovery of the equation for blackbody radiation built on the empirical law discovered by Wilhelm Wien, a less-known but still prominent physicist, who himself built his ideas on the work of the little-known English physicist George Searle.\cite{Wien1900} In this way, progress in science is more an accumulation of small efforts than a set of distinct ``heroic'' leaps in our understanding. 

Further weakening the wall separating the great from the rank-and-file, we can examine two examples of great innovations which made an individual researcher famous, but which were actually the cumulative effort of a team of scientists and engineers, most members of which are little remembered. The first example is Mie scattering, a theory that is used throughout physics for calculating the distribution of light scattered from spherically symmetric particles larger than the light wavelength, such as for water droplets in air. The German mathematician Alfred Clebsch (1861) was the first to deal with this problem, and worked out much of its mathematical details, but these were not widely recognized as important at the time. The Danish theorist Ludvig Lorenz (1890) later published a monumental memoir on the subject, building on the work of Clebsch, but published in a little-read journal and died soon afterwards, so that his work never received attention.\cite{Logan1965} The next 50 years were to see a series of famous names independently produce the results which can be found in Lorenz' long manuscript. The next to deal with the subject were J.\ W.\ Nicholson (1906), Peter Debye (1908), and Gustav Mie (1908) himself. Rather than deriving a series of new fundamental results, Mie's innovation was primarily to express the equations in a form more suitable to numerical calculations, and to assume that the optical constants of bulk metal materials can be used to represent optical properties of metal nanoparticles as well.\cite{Mie1908} This latter assumption allowed him to apply the theory to an example of contemporary interest: the colors produced by colloids of gold particles, but it was the former innovation that was crucial for associating his name with the theory. In fact, Mie's paper was not recognized as foundational by the scientific community at the time, since its application required prodigious calculations. Even Mie himself did not seem to give it much weight, and it was only with the advent of computers in the 1950s that the computations could be made practical.\cite{Horvath2009} By that time, the theory had been refined by another series of researchers that worked to generalize the approach so that it could be applied to a wider realm of scattering problems: T.\ J.\ I'A.\ Bromwich (1920), G.\ N.\ Watson (1918), F.\ P.\ White (1922), B.\ van der Pol (1937), and H.\ Bremmer (1937) all contributed towards advancing the theory and making it into the useful tool so widely applied today.\cite{Logan1965} The decision to attach Mie's name to the overall construct thus occurred long after the publication, by selecting his paper as the first to frame the problem in a way suited to computer calculation.

A second example of scientific teamwork is the Kalman filter --- a signal processing technique named for Rudolf E.\ K{\'a}lm{\'a}n, though Peter Swerling simultaneously developed a similar algorithm. Stanley Schmidt is generally credited with developing the first implementation of a Kalman filter. He --- or a member of his team of engineers --- made the important practical realization that the filter could be divided into two distinct parts, with one part for time periods between sensor outputs and another part for incorporating measurements, dramatically improving its utility. Later he and his group also realized that modifying the problem to linearize about the estimated state rather than the current state further improves performance, and is now referred to as the ``extended Kalman filter''.\cite{Grewal2010} James Potter next found a way to improve the algorithm's numerical stability through the procedure now known as square-root filtering. Finally, Richard S. Bucy also contributed to the theory, showing how to convert its use for continuous functions.\cite{Kailath1974} Thus, while Kalman's initial concept was the starting point, it required a team of scientists to adapt it and improve upon it in order for it to become the important tool that it is today.


These are two important discoveries where the name attached to them did only a secondary part of the effort, and sequence of smaller discoveries by a team of researchers built upon the initial theoretical concept to make it both more practical and more general. In this way, discoveries which immortalize a scientist's name are often derived from the work of a team, and can arise out of often-disdained technical drudgery of small improvements. Thus, it can be surprisingly difficult to appraise the quality of a scientist's oeuvre and decide on his or her greatness until long past their time, and even then we find that many discoveries are misattributed to someone else.\cite{Jackson2008} Implicit in this conclusion is that there must be many scientists who were regarded as of the 1st rank during their lifetimes but who were later downgraded to merely ``good''. And indeed there are many, though it requires a close reading of the contemporary scientific literature to find them. 

\section{Deep discoveries and shallow drudgery}

Much of the conversation among scientists centers around praise for great achievements and indirectly derogates ordinary science as drudgery. Yet in our daily lives as scientists, even for the greatest among us, it is in combating these ordinary tasks that we spend most of our time. And it is often this drudgery that fuels the insights needed to push further into the unknown. Tycho Brahe's time-consuming toil of building precision astronomical data was necessary for Johannes Kepler to found a new theory of planetary orbits. Kepler himself condemned as a ``treadmill of mathematical calculations'' the laborious work needed to prove his most famous results, while constantly desiring to return to his ``philosophical speculations''.\cite{Liscia2018} Yet it is for the insights derived from his tedious labors that he is still known today, while his philosophical speculations are generally regarded as uninteresting. All research involves tedium. While some of the tedium leads nowhere much, some of it also leads to important results, and it is only in hindsight that we can see the difference.

From the examples of achievements given lesser recognition, we can see that what scientists regard as the defining difference between the great and the merely good is \emph{profundity} much more than productivity. A productive engineer is treated with a degree of grudging respect, but little more than that, while a profound new theory is treated with a degree of awe. The following quote from a theoretical physicist is a typical outgrowth of this perspective:~\cite{Rao2015}
\begin{quotation} 
   {\noindent}Many people can solve a problem. But the [important] skill is to come up with ... the right problems. [Problems] that are not going to be only incremental progress but [which] really could make a difference. ...That is what distinguishes great scientists from good scientists --- the ability to [see] what is really worth working on.
\end{quotation}
Many researchers have been given similar advice. In a series of interviews with Nobel laureates, Robert Merton concluded that ``Almost to a man [Nobel laureates have in interviews laid] great emphasis on the importance of problem-finding, not only problem-solving. They [emphasize the need to seize] upon problems that are of fundamental importance [in contrast to the pedestrian work involving] endless detail or work just to improve accuracy....''\cite{Merton1968} And yet Nobelists have also been notoriously bad at doing this. Constant striving for profundity can often lead to failure, stifling progress and turning a scientist's attention away from what is curious to what is ``deep''. This trap has caught many physicists attracted by their ambitions, ignoring the fact that little problems sometimes turn out to be important pieces of bigger problems, and which by luck may turn out to be pieces of deep puzzles. 

This behavior of working with an eye towards achieving recognition also underlies the vexing behavior of modern science where researchers hop from one trendy research topic to another, as the pulse of the crowd changes from year to year. Young researchers seeking the acclaim of their peers are forced to focus on creating a research program that will be \emph{influential} rather than what is scientifically interesting. While it is hard to disagree that this is good advice for a successful career in modern science, it illustrates the degree to which science has become a profession rather than an avocation. Using the hierarchical language reflexively used by many physicists, perhaps we're all engineers now, even those who vehemently protest otherwise.

Despite our yearning for success on deep problems, scientific work involves much tedium and care towards wearisome detail, so that our daily lives contrast with our ambitious goals. However, if see this ``problem of greatness'' in the eye that we view our own publications, then much of the tension between ambition and reality falls away. Scientific publications show a polished version of actual events, displaying the successes and only rarely the failures encountered to reach the conclusions. While this polishing process has been maligned by many as a kind of dissembling, we can also recognize it as being inevitable. Anyone who has read through another person's scientific journals likely understands that there is little that is more tedious and uninteresting. While reading we have to suppress the constant urge to flip through the pages and get to the conclusion. This is exactly what the polished publication does for us. In the same way, when a scientific advance achieves recognition for being important, the polished version of it is what survives in peoples minds, and the messy work of getting there is forgotten as uninteresting. But that is a different thing from saying that it is unimportant; the tedium is necessary to build the final result. We may forget the little achievements needed to get there, but the result would have been impossible without them.


\section{Why focus on priority?}


After the long look into what makes an elite and what makes an ordinary scientist, a reader may come to feel that the constant focus on priority for discoveries seems senseless. Whether Newton or Leibniz, or for that matter some unknown John Smith, invented calculus has no real bearing on the working of science today. Establishing priority and celebrating discovery may be the currency by which scientists are rewarded, but if someone else independently discovered the same thing a year later, then it is hard to exalt the original advance as fundamental and unique.\cite{Zuidervaart2010} Instead, such discoveries have the feeling of being inevitable in the long run, so that the timing of the original discovery is contingent on a lucky confluence of conditions. Modern science in particular is pervaded with these discoveries. Scientific publications these days are all reports of small advances, not big leaps, and build heavily upon the prior work of other researchers, so that the interconnectedness of the whole process often makes it difficult to attribute primacy to a single individual or a small group.

Viewing the sociology of science from a distance in this way, it becomes hard to take seriously the innumerable squabbles over priority, especially when the antagonists protest their dedication to the purity of scientific pursuit and indifference towards compensation. Thankfully, not everyone joins in the bickering. Peter Fellgett, after looking back over his career, reflected that~\cite{Fellgett1984}
\begin{quotation}
   {\noindent}the greatest pleasure ... comes from recalling how, when we were all striving to create something new, there were from time to time inevitably disagreements and tensions, yet when we meet today these have been purged by time and experience and ... [we see how the] ideas originating from each of us were plied and woven into ... the tapestry of human achievement.
\end{quotation}
Thus, many scientists do see these quarrels over priority as unseemly, and have shown ambivalence towards pushing priority on their own discoveries. Outright dismissing of priority may seem an option, but is not one that the scientific community has been willing to accept as it disaffirms the value of originality and promotes wasteful repetition. This tension between insisting on valuing priority versus recognizing its unseemliness forces on each individual a difficult choice over how to deal with the conflicting norms.\cite{Merton1957}


\section{Conclusion}


For someone who focuses on immortalizing his or her achievements into the history of science, failing to achieve greatness is a problem of unfulfilled ambition for personal gain. However, for a scientist who learns to focus instead on the joy of learning and discovery, and leave the disputes over priority to others, then the labor is its own reward. For those who adhere to the latter view, the annual outcry over the awarding of the Nobel prize is a pointless waste. Whether a scientist is widely recognized enough to be awarded such a high-profile prize should not be a reflection on whether or not their work is valuable. Peter Fellgett, in response to hearing one such complaint, replied that ``It may be true that some laboratories make the `pursuit of Nobel prizes [as] the ultimate aim'. If so, the more fools they [are].... One expects scientists to have something better to do than to display the envy of socialites denied admittance to an exclusive club.''\cite{Fellgett1975}

In modern science, the faster pace of communication means that the piecemeal advances of science are becoming steadily smaller and more numerous. Scientific publications exhibit smaller research advances than before, and the work is increasingly distributed among team members than a solitary effort. As a result, the ``greats'' of today have less leisure to let their theories mature before deciding to publish. There are more mistakes, more partial results, more knitting of results among multiple researchers. The time of great discoveries by individuals has largely passed. And perhaps the scientists left unrecognized are more important than science generally credits us to be.



\small


\end{multicols}

\end{document}